\documentclass[a4paper]{jpconf}
\usepackage{graphicx}
\usepackage{multicol}

\begin{document}
\title{Investigating the correlations of flow harmonics in 2.76A TeV Pb--Pb collisions}

\author{Xiangrong Zhu$^{1,2,3}$, You Zhou$^{4}$, Haojie Xu$^{2,3}$, Huichao Song$^{2,3,5}$}
\address{$^1$School of Science, Huzhou University, Huzhou 313000, China}
\address{$^{2}$Department of Physics and State Key Laboratory of Nuclear Physics and Technology, Peking University, Beijing 100871, China}
\address{$^{3}$Collaborative Innovation Center of Quantum Matter, Beijing 100871, China}
\address{$^{4}$Niels Bohr Institute, University of Copenhagen, Blegdamsvej 17, 2100 Copenhagen, Denmark}
\address{$^{5}$Center for High Energy Physics, Peking University, Beijing 100871, China}
\ead{Huichaosong@pku.edu.cn}

\begin{abstract}
This proceeding briefly summarizes our recent investigations on the correlations of flow harmonics in 2.76A TeV Pb--Pb collisions with viscous hydrodynamics {\tt VISH2+1}. We calculated both the symmetric cumulants $SC^{v}(m, n)$ and the normalized symmetric cumulants $NSC^{v}(m, n)$, and found $v_{2}$ and $v_{4}$, $v_{2}$ and $v_{5}$, $v_{3}$ and $v_{5}$ are correlated, $v_{2}$ and $v_{3}$, $v_{3}$ and $v_{4}$ are anti-correlated. We also found $NSC^{v}(3, 2)$ are insensitive to the QGP viscosity, which are mainly determined by the initial conditions. 
\end{abstract}

\section{Introduction\label{sec:intro}}
The ultra-relativistic heavy-ion collision programs at RHIC and LHC have been utilized to the produce extreme conditions to create and study the strongly interacting Quark-Gluon Plasma (QGP), a deconfined state of quarks and gluons. One of the observables to probe the properties of the hot QCD matter is the azimuthal anisotropy in the momentum distribution of the produced particles. The anisotropic flow coefficient $V_{n}$ is generally defined through a Fourier decomposition of the emitted particle distribution as a function of the azimuthal angle $\varphi$, $P(\varphi) = \frac{1}{2\pi} \sum_{n=-\infty}^{+\infty} {\overrightarrow{V_{n}} \, e^{-in\varphi} }$ where $\overrightarrow{V_{n}} =v_{n}\,e^{in\Psi_{n}}$. The $v_{n}$ is the $n$-th order anisotropic flow harmonics and $\Psi_{n}$ is the symmetry plane angle. Recently, the correlations between different order $\overrightarrow{V_{m}}$ and $\overrightarrow{V_{n}}$ have been investigated both theoretically and experimentally, which not only focus on the correlations of the orientations of different flow-vector $\Psi_{n}$~\cite{Aad:2014fla,Qiu:2012uy,Teaney:2012gu,Jia:2012ju,Niemi:2015qia} but also on the correlations of the magnitudes of different flow-vector $v_{n}$~\cite{ALICE:2016kpq,Aad:2015lwa,Niemi:2012aj, Giacalone:2016afq,Qian:2016pau,Zhu:2016puf}.

In this proceeding, we will briefly review our recent investigations on the correlations of flow harmonics in 2.76A TeV Pb--Pb collisions using the event-by-event viscous hydrodynamics {\tt VISH2+1} with different initial conditions and  the QGP shear viscosity~\cite{Zhu:2016puf}.
 
\section{Setup of the calculation\label{sec:setup}}
The {\tt VISH2+1} is a (2+1)-d viscous hydrodynamic model to describe the fluid expansion of the QGP with longitudinal boost-invariance~\cite{Song:2007fn, Shen:2014vra}. In the following calculations, we use an equation of state (EoS) {\tt s95p-PCE}~\cite{Huovinen:2009yb}, which matches the partially chemical equilibrium hadron resonance gas at low temperature and the lattice QCD data at high temperature. Three different initial conditions, {\tt MC-Glauber}, {\tt MC-KLN}~\cite{Drescher:2006pi,Hirano:2009ah}, and {\tt AMPT}~\cite{Xu:2016hmp}, are used in our calculations to study the influence of initial conditions on the correlations of flow harmonics. To explore the sensitivity of the QGP shear viscosity, we choose two values of the specific shear viscosity $\eta/s$ for each initial condition. More specifically, $\eta/s=$ 0.08 and 0.20, for the {\tt MC-Glauber} and {\tt MC-KLN} initial conditions, and $\eta/s=$0.08 and 0.16 for the {\tt AMPT} initial conditions. The hydrodynamic output is converted to final hadron distributions along the freeze-out surface at the temperature $T_{dec}$ = 120 MeV via the Cooper-Frye prescription~\cite{Shen:2011eg,Song:2013qma}. 
The initial time of hydrodynamic evolution $\tau_{0}$ and the normalization factors of initial entropy density profiles have been tuned to fit the 0-5\% centrality data of $dN/d\eta$ and $p_{\rm T}$ spectra of $\pi$, $K$, and $p$.  
The bulk viscosity, net baryon density, and the heat conductivity are set to zero to simplify the calculations. 

\section{Results and discussion\label{sec:results}}
\begin{figure*}[ht]
  \centering
  \includegraphics[width=0.495\linewidth, height=5cm]{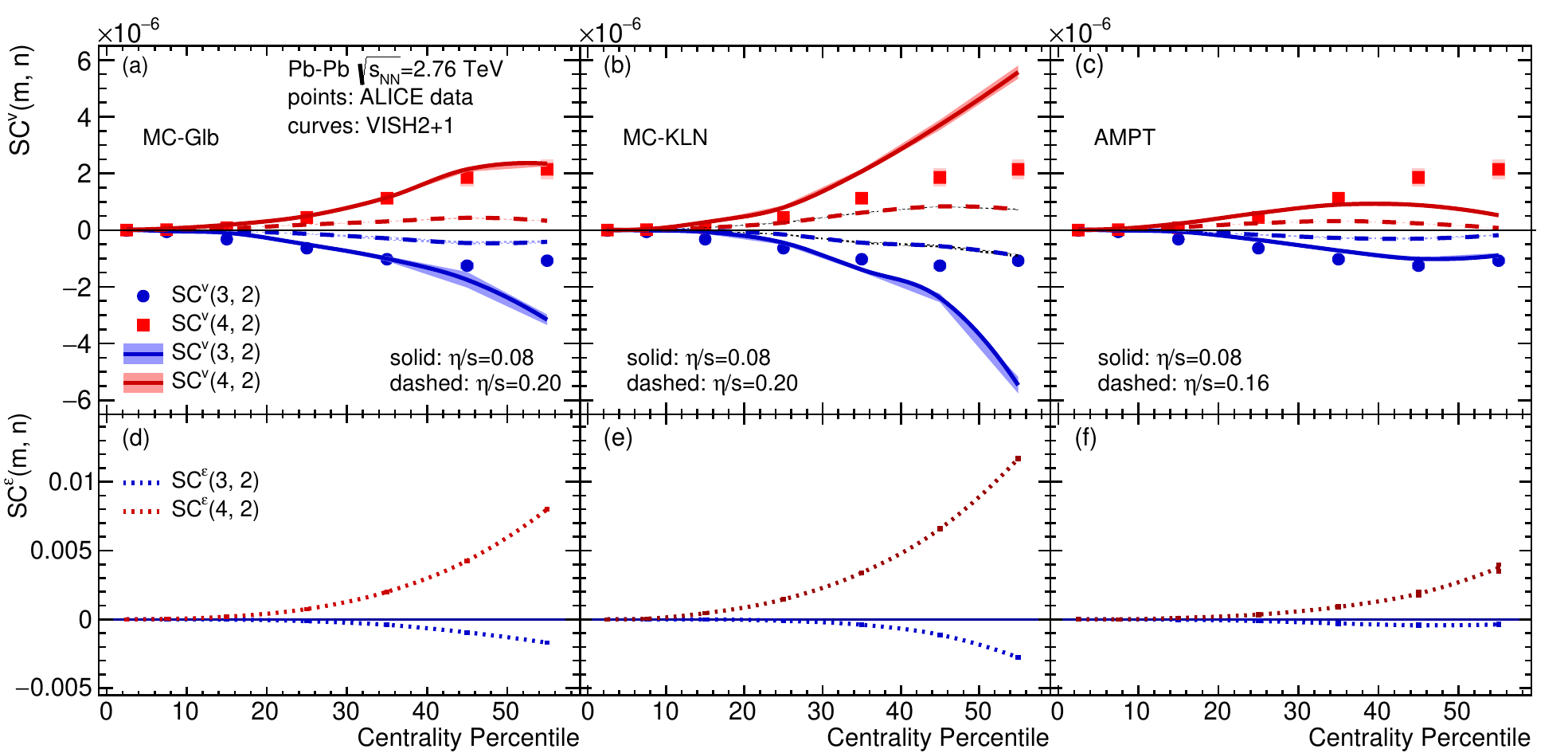}
  \includegraphics[width=0.495\linewidth, height=5cm]{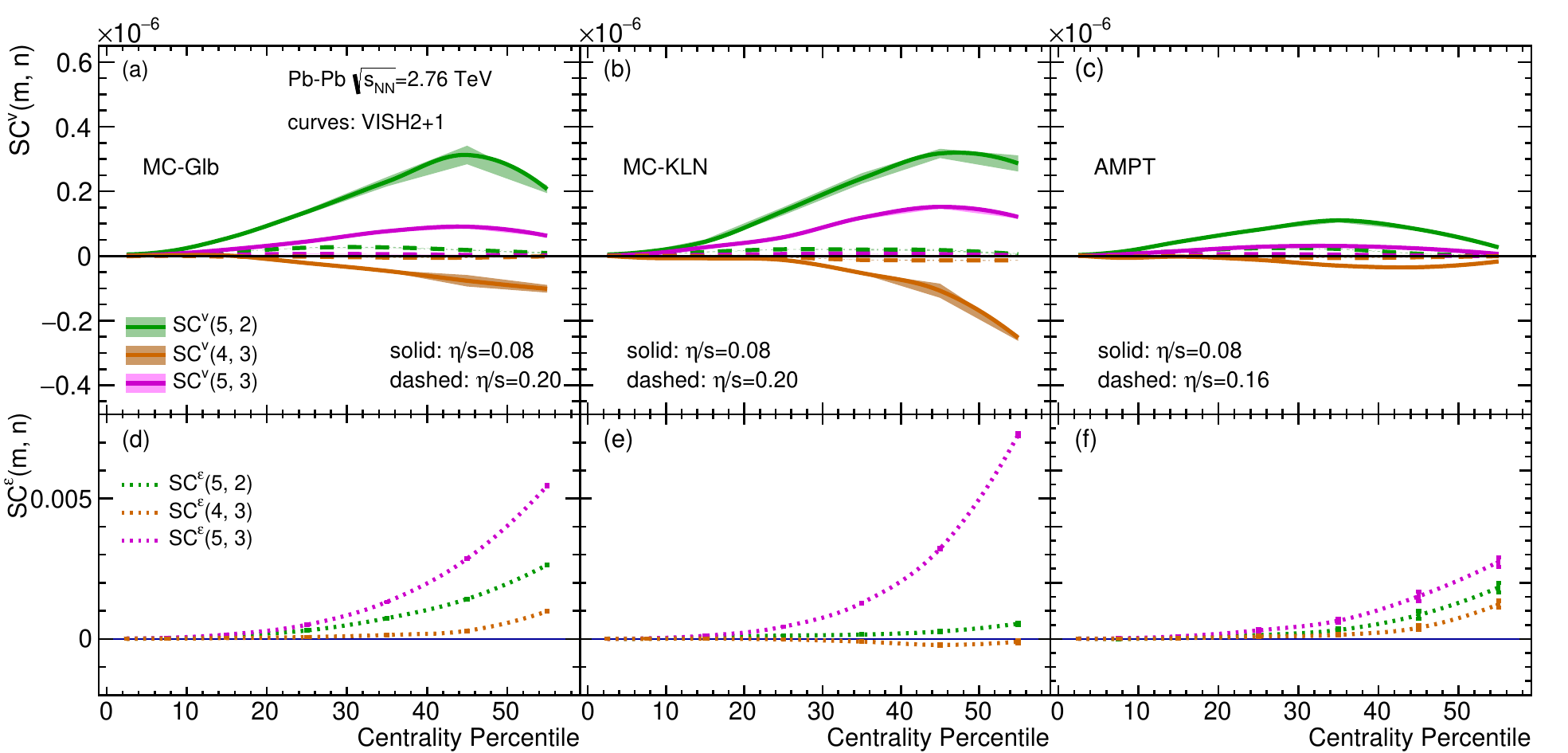}
\caption{(Color online) Left: Symmetric cumulants $SC^{v}(3, 2)$ and $SC^{v}(4, 2)$, and the corresponding symmetric cumulants for the initial state $SC^{\varepsilon}(m, n)$ in 2.76A TeV Pb--Pb collisions. The ALICE measurements are taken from~\cite{ALICE:2016kpq}. Right: Predicted symmetric cumulants $SC^{v}(5, 2)$, $SC^{v}(5, 3)$, and $SC^{v}(4, 3)$ in 2.76A TeV Pb--Pb collisions, together with their corresponding symmetric cumulants from the initial state.
\label{fig:scvmvn1}}
\end{figure*}

We firstly calculate the symmetric cumulants $SC^{v}(m, n)$ defined as $SC^{v}(m, n)= \left< v_{m}^{2} \, v_{n}^{2} \right> - \left< v_{m}^{2} \right> \left< v_{n}^{2} \right>$. The upper panels in Fig.~\ref{fig:scvmvn1} (left) show the comparison between our calculations and the ALICE measurements. We find that, for these initial conditions and different values of $\eta/s$, {\tt VISH2+1} calculations qualitatively capture the centrality dependence of the flow correlations, but not quantitatively. Specially, even though {\tt VISH2+1} with {\tt AMPT} initial conditions gives good descriptions for the integrated flow $v_{n}$ ($n\leq 4$)~\cite{Zhu:2016puf}, it can only reproduce the typical features of the correlations of flow harmonics. This indicates that the correlations between different flow harmonics are more sensitive to the details of hydrodynamic calculations than the individual $v_{n}$ coefficients alone.   

Similar to the ALICE data, our model gives negative $SC^{v}(3, 2)$ and positive $SC^{v}(4, 2)$, which suggests $v_2$ and $v_3$ are anti-correlated, while $v_2$ and $v_4$ are correlated. The results reveal that, for a given event, the case with an elliptic flow $v_{2}$ larger than the averaged $\langle v_{2} \rangle$ enhances the probability of finding a triangular flow $v_{3}$ smaller than the averaged $\langle v_{3} \rangle$ and the probability of finding a quadrangular flow $v_{4}$ larger than the averaged $\langle v_{4} \rangle$. The strengths of $SC^{v}(3, 2)$ and $SC^{v}(4, 2)$ are more suppressed with larger $\eta/s$ for each initial condition, which suggests that both $SC^{v}(3, 2)$ and $SC^{v}(4, 2)$ are strongly influenced by the QGP viscosity. By comparing with the symmetric cumulants of the initial state, $SC^{\varepsilon}(m, n)$, we observe the signs of $SC^{v}(3, 2)$ and $SC^{v}(4, 2)$ are determined by the signs of $SC^{\varepsilon}(3, 2)$ and $SC^{\varepsilon}(4, 3)$, respectively. 

Figure~\ref{fig:scvmvn1} (right) presents our predictions for the centrality dependent $SC^{v}(m, n)$ with $(m, n)=$ (5, 2), (5, 3), and (4, 3), together with their corresponding correlators $SC^{\varepsilon}(m, n)$ from the initial state. 
We observe that, for each initial condition, the {\tt VISH2+1} gives positive values for $SC^{v}(5, 2)$ and $SC^{v}(5, 3)$, and negative values for $SC^{v}(4, 3)$. This reveals $v_{2}$ and $v_{5}$, $v_{3}$ and $v_{5}$ are correlated, while $v_{3}$ and $v_{4}$ are anti-correlated. 
We also notice that their correlation strengths become weaker with the increase of $\eta/s$. The signs of $SC^{v}(5, 2)$ and $SC^{v}(5, 3)$ are consistent with their initial state correlators $SC^{\varepsilon}(5, 2)$ and $SC^{\varepsilon}(5, 3)$. However, $SC^{v}(4, 3)$ and $SC^{\varepsilon}(4, 3)$ show opposite signs for the {\tt MC-Glauber} and {\tt AMPT} initial conditions. This can be well understood from the proposed relationship of $v_{4}e^{i4\Phi}=a_{0}\varepsilon_{4}e^{i4\Psi_{4}}+a_{1}(\varepsilon_{2}e^{i2\Psi_{2}})^{2}$~\cite{Gardim:2011xv,Teaney:2012ke}, where the $\varepsilon_2^2$ term makes the dominant contributions in non-central collisions~\cite{Yan:2015jma}. As a result, the signs of $SC^{v}(4, 3)$ are affected by the correlation between $\varepsilon_{2}$ and $\varepsilon_{3}$ and the correlation between $\varepsilon_{3}$ and $\varepsilon_{4}$ rather than the correlation between $\varepsilon_{3}$ and $\varepsilon_{4}$ alone. 

\begin{figure*}[ht]
\centering
 \includegraphics[width=0.8\linewidth, height=8cm]{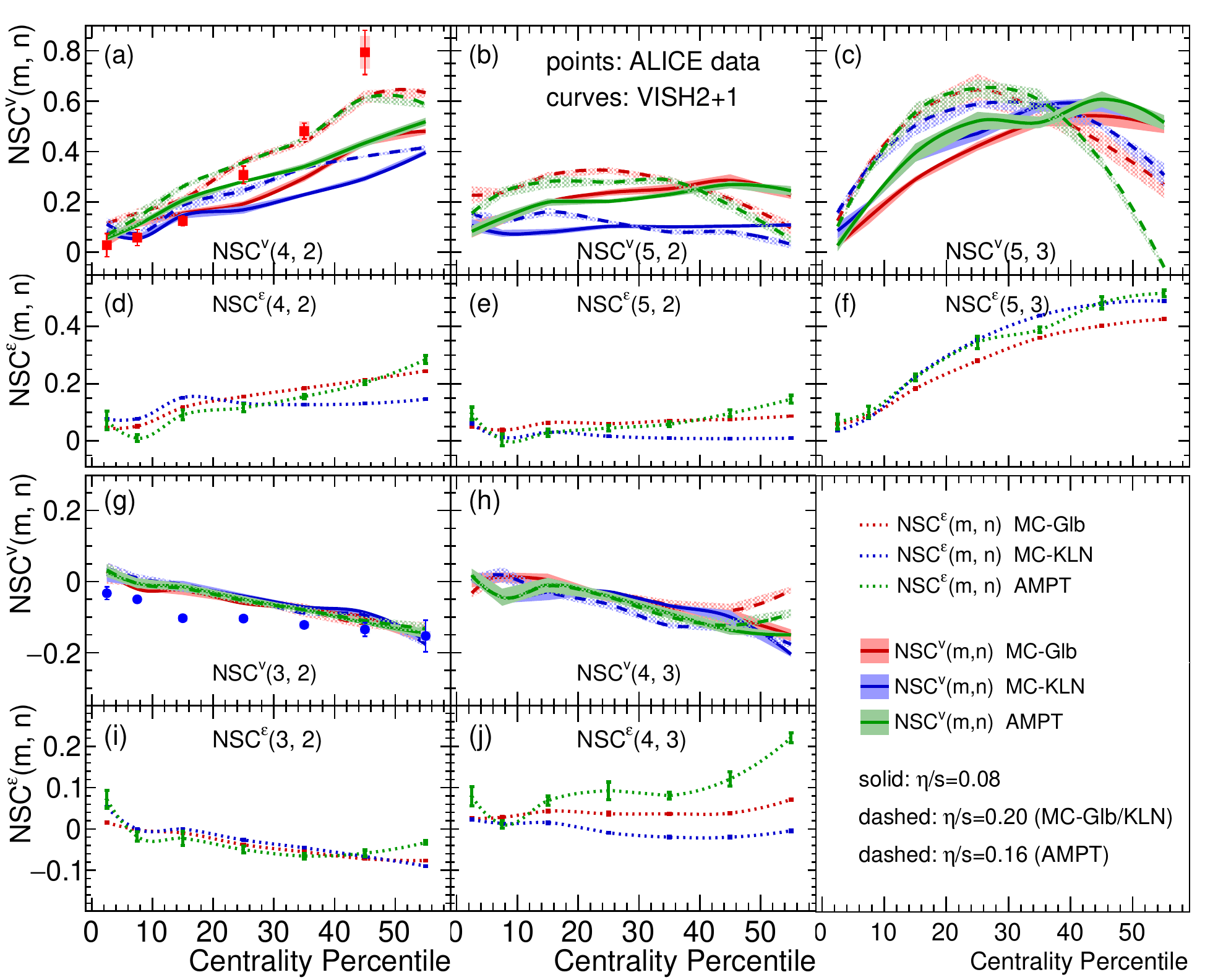}  \caption{(Color online) Normalized symmetric cumulants $NSC^{v}(m, n)$ and normalized symmetric cumulants of the initial state $NSC^{\varepsilon}(m, n)$  in 2.76A TeV Pb--Pb collisions.\label{fig:scRvmvn}}
\end{figure*}
Figure~\ref{fig:scRvmvn} shows the normalized correlator of flow harmonics and initial eccentricity coefficients, which are defined as $NSC^{v}(m, n)=SC^{v}(m, n)/\langle v_{m}^2 \rangle\langle v_{n}^2 \rangle$ and $NSC^{\varepsilon}(m, n)=SC^{\varepsilon}(m, n)/\langle \varepsilon_{m}^2 \rangle\langle \varepsilon_{n}^2 \rangle$, respectively. We find that $NSC^{v}(4, 2)$, $NSC^{v}(5, 2)$, and $NSC^{v}(5, 3)$ are sensitive to both initial conditions and $\eta/s$. Meanwhile, their corresponding $NSC^{\varepsilon}$ correlator are separated for different initial conditions. Compared to the ALICE data~\cite{ALICE:2016kpq}, the calculated $NSC^{v}(4, 2)$ are roughly fit the data for {\tt AMPT} initial conditions and $\eta/s=0.16$ and for {\tt MC-Glauber} initial conditions and $\eta/s=0.2$. This indicates the normalized symmetric cumulants can be used to constrain the QGP viscosity for different initial conditions. 
The $NSC^{v}(3, 2)$ from different combinations of initial conditions and $\eta/s$ are all roughly fit the ALICE data, which also roughly overlap with each other. Such $\eta/s$ independent character of $NSC^{v}(3, 2)$ can be naturally understood from the widely accepted results $v_2\approx k_{1}\varepsilon_2$ and $v_3\approx k_{2}\varepsilon_3$, where $k_{1}$ and $k_{2}$ are the proportion coefficients. Meanwhile, the $NSC^{\varepsilon}(3, 2)$ from the three initial conditions used in our calculations also almost overlap from central to semi-central collisions. In contract, panels (h) and (j) show that, although the $NSC^{\varepsilon}(4, 3)$ strongly depends on the initial conditions, the $NSC^{v}(4, 3)$ almost overlap, which is insensitive to the initial conditions used in our calculation. 

\section{Summary\label{sec:summary}}
In summary, we investigated the correlations between flow harmonics in 2.76A TeV Pb--Pb collisions using the event-by-event viscous hydrodynamics {\tt VISH2+1} with {\tt MC-Glauber}, {\tt MC-KLN}, and {\tt AMPT} initial conditions. We found the symmetric cumulants $SC^{v}(m, n)$ are sensitive to both initial conditions and the QGP shear viscosity, The normalized symmetric cumulants $NSC^{v}(3, 2)$ are mainly determined by the correlation in the initial state, which are insensitive to the QGP viscocity.
In contrast, $NSC^{v}(4, 2)$, $NSC^{v}(5, 2)$, $NSC^{v}(5, 3)$ are sensitive to both initial conditions and $\eta/s$. We found that the correlations of flow harmonics are more sensitive to the details of theoretical calculations than individual flow harmonics, which could be used for further constraint the properties of the QGP.

\ack
This work is supported by the NSFC and the MOST under grant Nos.11435001 and 2015CB856900, and partially supported by China Postdoctoral Science Foundation under grant No. 2015M570878 and 2015M580908, by the Danish Council for Independent Research, Natural Sciences, and the Danish National Research Foundation (Danmarks Grundforskningsfond). 


\section*{References}
\begin{thebibliography}{9}

\bibitem{Aad:2014fla} 
  G.~Aad {\it et al.} [ATLAS Collaboration],
  Phys.\ Rev.\ C {\bf 90}, no. 2, 024905 (2014).

\bibitem{Qiu:2012uy}
  Z.~Qiu and U.~Heinz,
  Phys.\ Lett.\ B {\bf 717}, 261 (2012).

\bibitem{Teaney:2012gu}
  D.~Teaney and L.~Yan,
  Nucl.\ Phys.\ A {\bf 904-905}, 365c (2013).

 \bibitem{Jia:2012ju}
  J.~Jia and D.~Teaney,
  Eur.\ Phys.\ J.\ C {\bf 73}, 2558 (2013).

\bibitem{Niemi:2015qia}
 H.~Niemi, K.~J.~Eskola and R.~Paatelainen,
 Phys.\ Rev.\ C {\bf 93}, no. 2, 024907 (2016).


\bibitem{ALICE:2016kpq}
  J.~Adam {\it et al.} [ALICE Collaboration],
  arXiv:1604.07663 [nucl-ex];

\bibitem{Aad:2015lwa}
  G.~Aad {\it et al.} [ATLAS Collaboration],
  Phys.\ Rev.\ C {\bf 92}, no. 3, 034903 (2015).
\bibitem{Niemi:2012aj}
  H.~Niemi, G.~S.~Denicol, H.~Holopainen and P.~Huovinen,
  Phys.\ Rev.\ C {\bf 87}, no. 5, 054901 (2013).

\bibitem{Giacalone:2016afq}
  G.~Giacalone, L.~Yan, J.~Noronha-Hostler and J.~Y.~Ollitrault,
  arXiv:1605.08303 [nucl-th].

\bibitem{Qian:2016pau}
  J.~Qian and U.~Heinz,
  arXiv:1607.01732 [nucl-th].

\bibitem{Zhu:2016puf}
  X.~Zhu, Y.~Zhou, H.~Xu and H.~Song,
  arXiv:1608.05305 [nucl-th].

\bibitem{Song:2007fn}
  H.~Song and U.~Heinz,
  Phys.\ Lett.\  {\bf B658}, 279 (2008);
  Phys.\ Rev.\  C {\bf 77}, 064901 (2008);
  Phys.\ Rev.\ C {\bf 78}, 024902 (2008);
  H.~Song, Ph.D Thesis, The Ohio State University, August 2009,
  arXiv:0908.3656 [nucl-th].

\bibitem{Shen:2014vra}
  C.~Shen, Z.~Qiu, H.~Song, J.~Bernhard, S.~Bass and U.~Heinz,
  Comput.\ Phys.\ Commun.\  {\bf 199}, 61 (2016).

\bibitem{Huovinen:2009yb}
  P.~Huovinen and P.~Petreczky,
  Nucl.\ Phys.\  {\bf A837}, 26 (2010);
  C.~Shen, U.~Heinz, P.~Huovinen and H.~Song,
  Phys.\ Rev.\ C {\bf 82}, 054904 (2010).

\bibitem{Drescher:2006pi}
  A.~Adil, H.~J.~Drescher, A.~Dumitru, A.~Hayashigaki and Y.~Nara,
  Phys.\ Rev.\  C {\bf 74} 044905 (2006);
  H.~J.~Drescher and Y.~Nara,
  {\it ibid.} {\bf 76} 041903 (2007).

\bibitem{Hirano:2009ah}
  T.~Hirano and Y.~Nara,
  Phys.\ Rev.\  C {\bf 79} 064904 (2009);
  and
  Nucl.\ Phys.\ {\bf A830} 191c (2009).

\bibitem{Xu:2016hmp} 
  H.~j.~Xu, Z.~Li and H.~Song,
  Phys.\ Rev.\ C {\bf 93}, no. 6, 064905 (2016).

\bibitem{Shen:2011eg} 
  C.~Shen, U.~Heinz, P.~Huovinen and H.~Song,
  Phys.\ Rev.\ C {\bf 84}, 044903 (2011).

\bibitem{Song:2013qma} 
  H.~Song, S.~Bass and U.~W.~Heinz,
  Phys.\ Rev.\ C {\bf 89}, no. 3, 034919 (2014);
  X.~Zhu, F.~Meng, H.~Song and Y.~X.~Liu,
  Phys.\ Rev.\ C {\bf 91}, no. 3, 034904 (2015).


\bibitem{Gardim:2011xv}
 F.~G.~Gardim, F.~Grassi, M.~Luzum and J.~Y.~Ollitrault,
    Phys.\ Rev.\ C {\bf 85}, 024908 (2012).

\bibitem{Teaney:2012ke}
  D.~Teaney and L.~Yan,
  Phys.\ Rev.\ C {\bf 86}, 044908 (2012).
\bibitem{Yan:2015jma}
  L.~Yan and J.~Y.~Ollitrault,
  Phys.\ Lett.\ B {\bf 744}, 82 (2015).


\end{thebibliography}
\end{document}